\documentclass{iaus}
\usepackage{graphicx}

\title[What will Gaia tell us about the Galactic disk?] 
{What will Gaia tell us about the Galactic disk?}

\author[C.A.L. Bailer-Jones]   
{Coryn A.L. Bailer-Jones}
\affiliation{Max-Planck-Institut f\"ur Astronomie, K\"onigstuhl 17,
69117 Heidelberg, Germany\\ email: {\tt calj@mpia.de} \\[\affilskip]}

\pubyear{2009}
\volume{254}  
\pagerange{???}
\date{June 2008}
\jname{The Galaxy disk in a cosmological context}
\editors{}
\begin{document}


\def\aabun{[$\alpha$/Fe]}
\def\av{A$_{\rm V}$}
\def\teff{${\rm T}_{\rm eff}$}
\def\logg{$\log g$}
\def\feh{[Fe/H]}
\def\age{$\tau$}
\def\parallax{$\pi$}

\def\deg{$^{\circ}$}
\def\micron{$\mu$m}
\def\uas{$\mu$as}

\def\Msol{M$_{\odot}$}
\def\Rsol{R$_{\odot}$}
\def\Lsol{L$_{\odot}$}


\maketitle

\begin{abstract}
  Gaia will provide parallaxes and proper motions with accuracy
  ranging from 10 to 1000 microarcsecond on up to one billion
  stars. Most of these will be disk stars: for an unreddened K giant
  at 6 kpc, it will measure the distance accurate to 15\% and the
  transverse velocity to an accuracy of about 1 km/s.  Gaia will
  observe tracers of Galactic structure, kinematics, star formation
  and chemical evolution across the whole HR diagram, including
  Cepheids, RR Lyrae, white dwarfs, F dwarfs and HB stars. Onboard
  low resolution spectrophotometry will permit -- in addition to an
  effective temperature estimate -- dwarf/giant discrimination,
  metallicity measurement and extinction determination. For the first
  time, then, Gaia will provide us with a three-dimensional
  spatial/properties map and at least a two-dimensional velocity map
  of these tracers. (3D velocities will be obtained for the brighter
  sources from the onboard RV spectrograph). This will be a goldmine
  of information from which to learn about the origin and evolution of
  the Galactic disk.  I briefly review the Gaia mission, and then show
  how the expected astrometric accuracies translate into distance and
  velocity accuracies and statistics.  I then briefly examine the
  impact Gaia should have on a few scientific areas relevant to the
  Galactic disk, specifically disk structure and formation, the
  age--metallicity--velocity relation, the mass--luminosity relation,
  stellar clusters and spiral structure. Concerning spiral arms, I
  note how a better determination of their locations and pattern speed
  from their OB star population, plus a better reconstruction of the
  Sun's orbit over the past billion years (from integration through
  the Gaia-measured gravitational potential) will allow us to assess
  the possible role of spiral arm crossings in ice ages and mass
  extinctions on the Earth.

\vspace*{1em}\hspace{-1em}16 June 2008; corrected 8 September 2013 (K giant distance accuracy)

\end{abstract}

\firstsection 



\section{Gaia overview}
\vspace*{2ex}

Gaia is a high-accuracy astrometric satellite to be launched by ESA at
the end of 2011.  By measuring the positions of stars tens of times
over a five year baseline, it can derive the mean position, the
parallax and two-dimensional proper motion of each star. These
are five components of the six dimensional phase space, the
sixth component -- the radial velocity -- being provided for the
brighter stars by an onboard spectrograph.  Gaia delivers absolute
parallaxes tied to an inertial (quasar--based) reference frame. 
Gaia, which is currently under construction, represents a major step beyond the enormously successful Hipparcos
mission, and is currently the only large-scale astrometric mission beyond the planning stage.

Gaia will perform a survey of the entire sky complete to magnitude
G=20 (V=20--22). This covers some 10$^9$ stars, a million quasars and
a few million galaxies.  The sky coverage is complete to this
magnitude, bar about 1\% of the sky area where Gaia is confusion
limited.  Gaia will achieve an astrometric accuracy of 12--25\,\uas\
at G=15 (providing a distance accuracy of 1--2\% at 1 kpc) and
100--300\,\uas\ at G=20.  These numbers are also the approximate
parallax accuracy in \uas\ and the proper motion accuracy in
\uas/year.  The accuracy range reflects a colour dependency: better accuracy
is achieved for redder sources (as more photons are collected). Astrometry and photometry are done in
a broad (``white light'') band (G). Gaia will also measure radial
velocities to a precision of 1--15\,km/s for stars with V=17 via
R=11\,500 resolution spectroscopy around the CaII triplet (the
``Radial Velocity Spectrograph'', RVS).  To characterize all sources
(which are detected in real time), each is observed via low dispersion
prism spectrophotometry over 330--1000\,nm with a dispersion between 3
and 30 nm/pixel (the ``BP/RP'' instrument). From this we will estimate the ``usual''
astrophysical parameters, \teff, \logg\ and \feh, but also the
line-of-sight extinction to stars individually, plus perhaps also
[$\alpha$/Fe] in some cases.  Spectra from the radial velocity
spectrograph will help the parameter determination for some stars
brighter than about $V=14$, and will also allow the detection of
emission lines and abundance anomalies.

Gaia has a nominal mission duration of five years (plus a possible one
year extension), and 2--3 years following the end of operations are
planned to complete the data processing. (The astrometry is
self-calibrating, so the data must be reduced globally/simultaneously
to get the final solutions and best astrometric accuracy.) The
final catalogue will be available in about 2020, proceeded by earlier data releases. For more
information on the satellite, science and data processing see {\tt
  http://www.rssd.esa.int/Gaia} and the proceedings volume by
\cite[Turon et al.\ (2005)]{turon05} (also available from the
website).

\section{Distance and velocity accuracy and statistics}
\vspace*{2ex}

Distances are vital in every area of astronomy. We need them to convert 2D
angular positions to 3D spatial coordinates, allowing us to reveal the
internal structure of stellar clusters or map the location of the spiral arms,
for example.  Knowing the distance we can convert 2D (angular) proper motions
to physical velocities, and the apparent luminosity to the intrinsic
luminosity, a fundamental quantity in stellar structure and evolution studies.
Parallaxes are the only method of direct distance determination
which does not make assumptions about the target source. We
can measure parallaxes to virtually anything (not just, say, eclipsing
binaries) and virtually all other rungs in the distance ladder are ultimately
calibrated by them.

\subsection{How accuracy varies with distance}

Apart from very bright sources, the parallax error from Gaia is
dominated by photon statistics.  Thus the parallax uncertainty,
$\sigma(\varpi)$, varies with the measured flux, $f$, as
$\sigma(\varpi) \propto 1/\sqrt{f}$. Therefore, for a source of given
intrinsic luminosity, the parallax error is linearly
proportional to distance, $\sigma(\varpi) \propto d$. The parallax
itself of course also decreases with increasing distance, $\varpi =
1/d$, so the fractional parallax error, $\sigma(\varpi)/\varpi$,
increases as distance squared,
$\sigma(\varpi)/\varpi \propto d^2$. Converting a parallax to a
distance is actually non-trivial, due to the need to account for
Lutz-Kelker biases (see \cite{bm98} for
a discussion). However, we can roughly equate this fractional parallax
error with the fractional distance error, hereafter abbreviated as
{\bf fde}.

The proper motion uncertainty varies with distance in the same way as
the parallax, and the measured proper motion itself, for given space
velocity, decreases linearly with increasing distance. Therefore, the
transverse velocity uncertainty (not fractional), also increases with
the square of the distance, $\sigma(v) \propto d^2$ (for a source with
fixed $M_V$). A useful relation to remember is that a proper motion of
1 mas/yr for a source at 1\,kpc corresponds to a transverse velocity
of about 5\,km/s.  (Note that this is by definition because the
parallax definition is related to the mean velocity of the Earth
around the Sun.)

\subsection{Accuracies for certain types of stars}

The above relations are useful, because they enable us to see to what
distance certain Galactic structure tracers can still be observed with
useful accuracy. Here are some examples. A K1 giant, with an intrinsic
luminosity of $M_V = 1.0$, will have an apparent magnitude of about
$G=14.5$ when at a distance of 6\,kpc (assuming zero interstellar
extinction). (For simplicity I ignore the colour dependence of both
the $V$ to $G$ relationship and the astrometric accuracy, adopting the
lower accuracy blue limit for the latter.) The fde is 15\% and
$\sigma(v)=0.75$\,km/s. This clearly allows us to trace this tracer
with high accuracy in phase space.

The astrometric accuracy, e.g.\ $\sigma(\varpi)$, is a factor of
$\sqrt{10}$ worse for a source which is 2.5\,mag fainter; similarly the
astrometric accuracy is a factor of two worse for a source which is
1.5\,mag fainter (equivalent to doubling the distance for a source of
constant $M_V$).

A G3 dwarf with $M_V = 5.0$ has $G=16.5$ at a distance of 2\,kpc, and
so a parallax accuracy of about 40\,\uas. The parallax itself is
500\,\uas, so the fde is 8\%. The transverse velocity accuracy is just
0.4\,km/s Moving it to 4\,kpc reduces the accuracy and parallax to
80\,\uas\ and 250\,\uas\ respectively, giving an fde of 30\%. 

The velocity accuracies in particular are very good when one considers
typical velocities within the Galaxy (e.g.\ the rotation of the LSR
about the Galactic centre is about 220\,km/s).  So as the distance
increases, it is parallaxes rather than the proper motions which first
become less useful, although at large distances the transverse
velocity uncertainty becomes dominated by the parallax error.  The
distance at which the parallax error is 50\% is 5\,kpc for the G3
dwarf (which then has $G=18.3$) and 13\,kpc for the K1 giant (which
then has $G=16.1$).  Thus we see that beyond some distance
which depends on intrinsic magnitude, the Gaia parallaxes
on {\em individual} objects are no longer much good, so we will have
to rely on spectroscopic parallaxes (derived from the onboard
spectrophotometry), in particular to convert proper motions to
velocities.

\subsection{Accuracy statistics}

While it's easy to throw out these numbers, we must not forget the
enormous improvement they represent in accuracy above what is
currently possible. The improvement with respect to Hipparcos comes primarily
through the larger field of view: Gaia simulaneously observes
tens of thousands of stars on a large CCD focal plane, whereas
Hipparcos could only observe one at a time (with a photomultiplier
tube, PMT) (\cite{lindegren05}). In addition, the larger QE detector
(ca. 75\% for CCDs compared to around 1\% for a PMT), the larger
collecting area (0.7\,$m^2$ compared to 0.03\,$m^2$) and the broader effective
bandpass together provide a much larger detectable photoelectron
flux per star.

The value of Gaia is not just accuracies of individual objects, but
also the fact that it is an all sky survey with a relatively deep
limiting magnitude (G=20), so these accuracies are achieved on large
numbers of objects. (Note that the bright -- saturation -- limit of
Gaia is $G=6$, so it can just reach stars visible to the naked eye.)
Roughly how many can be found via a Galaxy model.  We find that Gaia
will provide distances with an accuracy of 1\% or better for 11
million stars.  Taking 12\,\uas\ as the lower limit to the Gaia
parallax accuracy, this means that all of these stars must lie within
800\,\,pc (i.e.\ have a parallax less than 1250\,\uas).  Some 150
million stars will have a distance accuracy better than 10\% (all
within 8\,kpc) and some 100\,000 stars better than 0.1\% (all within
80\,pc). For comparison, currently fewer than 200 stars have had their
parallaxes measured to better than 1\%. All are within 10\,pc and all
were measured with Hipparcos.

\section{Astrophysical parameter estimation}
\vspace*{2ex}

Gaia does not use an input catalogue, so the properties of the sources
are not known a priori. Gaia only observes point sources and
the onboard processing generally bins the CCD data around a source in
one-dimension (so that only the higher spatial resolution along-scan
direction remains). Consequently there is little morphological information
(although some can be reconstructed from the multiple scans of a given
source which occur at a range of positions angles).  There are,
therefore, two main classification tasks with the Gaia data. The first
is a discrete source classification, to discriminate between stars,
galaxies and quasars (and perhaps also unresolved binaries and other
types). Identification of a clean quasar sample (\cite{bj08}) is
necessary in particular for the global astrometric solution.

The second step is to estimate the stellar astrophysical parameters
(APs). This is done primarily using the low resolution prism
spectrophotometry (BP/RP), as well as the parallax when available.
The classification software is still under development, but we have
preliminary estimates of the performance of a stellar parametrizer
based on support vector machine classifiers applied to synthetic
spectra convolved with a Gaia instrument and noise model. At $G=15$,
the RMS uncertanties in the derived APs are as follows: 1--5\% in
\teff\ for a wide range of effective temperatures (OB to K stars);
line-of-sight extinction, \av, to 0.05-0.1\,mag for hot stars;
$<0.2$\,dex in [Fe/H] for spectral types later than F down to
$-2.0$\,dex metallicity; \logg\ to between 0.1 and 0.4\,dex for all
luminosity classes, but to better than 0.1\,dex for hot star (OBA
stars). It should be noted that the performance varies a lot with the
physical parameters themselves, as well as with $G$ magnitude (i.e.\
signal-to-noise ratio). Work in this area is ongoing.

The inferred stellar parameters are tied, of course, to the stellar
models used to simulate the synthetic spectra on which the
classification models are trained. (If the models are instead trained
directly on real spectra, then these must somehow be parametrized using models, so
our results are always tied to physical stellar models). In addition to developing
classification algorithms, the Gaia DPAC (Data Processing and Analysis
Consortium) is also working on improving the stellar input physics and
model atmospheres. A ground-based observing programme will provide
some data to calibrate the fluxes of the models.

\section{Gaia science}
\vspace*{2ex}

I now outline a few areas of disk astrophysics where Gaia should make
significant contributions.

\subsection{Disk structure}

Many aspects of the thin and thick disk structure are not well known
(see \cite{vallenari05} for a summary). For example, thin disk scale
height estimates vary between 200 and 330\,pc. The inferred scale length varies
between 2--3\,kpc (as measured by star counts and integrated light)
and 2--4\,kpc (as measured by kinematics). The vertex deviation and
the vertial tilt are not well known.  All of these measurements of the
velocity elipsoids are based on relatively small samples. Gaia will
provide much larger samples with 3D spatial structure and 2D or 3D
velocity measures. Transverse velocities accurate to about 1\,km/s
will be achieved for some 50 million stars out to a few kpc.

\subsection{Substructure and mergers}\label{substructure}

One of the most important questions Gaia will address is that of how
and when the Galaxy formed. $\Lambda$CDM models of galaxy formation
predict that galaxies are built up by the hierarchical merger of
smaller components and some suggest that the halo is composed
primarily of merger remnants (for a review see \cite{freeman02}) .
From extragalactic observations there is good evidence for both the
accretion of small components and for the merging of similar-sized
galaxies. Within our own Galaxy, recent surveys -- 2MASS and SDSS in
particular -- have found the fossils of past and ongoing mergers in
the halo and possibly also the disk of our Galaxy. They have been
identified primarily as spatial overdensities in two-dimensional
(angular) photometric maps of large areas of the sky (sometimes
accompanied with radial velocities). In a few cases, distance measures
have been included by taking magnitude as a distance proxy
(\cite{belokurov06}) or by examining the 2D density of a limited range
of spectral types (i.e.\ using a spectroscopic parallax of some
tracers) (\cite{yanny00}).  But because merging satellites are
disrupted by the Galactic potential and the material spread out after
several orbits, density maps are a limited means of finding
substructure.  Without an accurate distance the interpretation of 2D
maps is plagued by projection effects.  Even with perfect 3D maps, the
contrast against the background (including other streams) is often low
(\cite{brown05}).  To improve this, we need 3D kinematics, i.e.\
radial velocities and proper motions (combined with distances). In an
axisymmetric potential the component of angular momentum parallel to
this axis ($L_z$) of a merging satellite is an integral of motion.  In
a static potential, the energy ($E$) is also an integral of motion
(\cite{bt87}).  Thus while a merging satellite could be well-mixed
spatially, it would remain (approximately) unperturbed in $(L_z, E)$ space.  Of
course, the Galactic potential is neither time-independent nor
perfectly axisymmetric and Gaia has measurement errors, but
simulations have demonstrated that Gaia will be able to detect
numerous streams 10\,Gyr or more (i.e.\ many orbits) after the start
of their disruption (\cite{helmi00}).

Gaia will perform a 5D phase space survey over the whole sky, with the sixth
component -- radial velocity -- being available for stars brighter
than V=17. At this magnitude, spectral types A5III, AOV and K1III (which all
have M$_V$ $\simeq 1.0$) are seen at a distance of 16\,kpc (for zero
extinction). The corresponding proper motion accuracy is about 50\,\uas/yr, or
4\,km/s. In addition to the
5D or 6D phase space information, Gaia provides abundances and ages for
individual stars. Search for patterns in this even higher dimensional space
permits an even more sensitive (or reliable) search for substructure.

\subsection{Dark matter}

Two distinct aspects of the Gaia mission permit us to study the mass and
distribution of dark matter in our Galaxy. First, from the 3D kinematics of
selected tracer stars, Gaia will map the total gravitational potential (dark
and bright) of our Galaxy, in particular the disk.  Second, from its
parallaxes and photometry, Gaia will make a detailed and accurate measurement
of the stellar luminosity function. This may be converted to a (present-day)
stellar mass function via a Mass--Luminosity relation (see
section~\ref{ML_relation}).  From this we can infer a stellar mass
distribution. Subtracting this from the total mass distribution obtained from
the kinematics yields the dark matter distribution.

\subsection{Stellar structure and stellar clusters}

Stellar luminosity is a fundamental property, so its
measurement across a range of masses, ages and abundances is a
critical ingredient for testing and improving stellar models. In open and
globular clusters an accurate determination of luminosities and effective
temperatures (which Gaia also provides) gives us the HR diagram for different
stellar populations. (To derive an accurate luminosity we also need an
accurate estimate of the line-of-sight extinction. This will be obtained
star-by-star from the Gaia spectrophotometry.) We may then address fundamental
questions of stellar structure, such as the bulk Helium abundance (which is
not observable in the spectrum), convective overshooting and diffusion.  One
of the main uncertainties in the age estimation in clusters is accurately
locating the main sequence (for open clusters) or main sequence turn off (for
globular clusters). Gaia's accurate parallaxes and unbiased
(magnitude-limited) survey will greatly improve this.

In addition to using clusters as samples for refining stellar
structure and evolution, we can also study them as populations. Gaia
will observe many hundreds of clusters, allowing us to determine the
(initial) mass function into the brown dwarf regime and examine its
dependence on parameters such as metallicity, stellar density and
environment. There are perhaps 70 open clusters and star formation
regions with 500pc. For these, Gaia will provide individual distances
to stars brighter than G=15 to better than 1\%. This will permit us,
for the first time, to map the 3D spatial structure of many clusters,
with a depth accuracy as good as 0.5--1\,pc for clusters at
200\,pc. From the 3D kinematics we can likewise study the internal
dynamics of a cluster. A G=15 star will have its proper motion
measured with an accuracy of 20\,\uas/yr, corresponding to a speed
uncertainty of 0.1\,km/s at this distance (half this for a red
star). The speed uncertainty varies linearly with the distance for a
fixed magnitude, so at 200\,pc the speed uncertainty is just
20\,m/s. With this accuracy we can measure the internal kinematics of
the cluster and so investigate the phenomena of mass segregation, low
mass star evaporation and the dispersion of clusters into the Galactic
field.  (I have assumed that the uncertainty in the transverse
velocity is dominated by the proper motion error and not the distance
error. For the sake of the proper motion to velocity conversion we can
assume all stars to be at the same distance, and therefore average
over many stars to get a more accurate distance.)

At larger distances Gaia will be able to say less about the internal
structure of an open cluster.  However, using the proper motions to
help define membership, the average distance to the cluster, plus its age
and metallicity, can be measured accurately. In this
way, we can use a few thousand open clusters over distances of tens of
kpc as tracers of the disk abundance gradient.

Just as Gaia should be able to identify the fossils of past mergers from
their phase space substructure (section~\ref{substructure}), so the 6D phase
space data plus astrophysical parameters for tens of millions of stars
will allow Gaia to detect new stellar clusters, associations or moving groups
based on their clustering in a suitable multi-dimensional parameter space. It
can likewise confirm or refute the existence of controversial clusters.

\subsection{Stellar mass--luminosity relation}\label{ML_relation}

Gaia will detect many binary systems. These are found in a number of
ways, as summarized by \cite{arenou05}. Most are found via the astrometry
as astrometric solutions which do not fit the standard 5-parameter
model. These include both Keplerian fits to the data (for periods up
to 10 years) and nonlinear motions (accelerated proper motions,
periods up to a few hundred years). Very wide binaries can be found as
common proper motion pairs. In addition, bright binaries can be found
with the radial velocity spectrograph and as eclipsing
binaries. Unresolved binaries of different spectral types
with brightness ratios not too far from unity can be found as part of
the classification work, from the identification of two spectral
energy distributions.

For those systems with orbital periods of about
ten years or less, Gaia can solve for the orbital elements and for the
total mass of the system. If the components of the system are
spatially resolved then we may determine their individual masses. Gaia
furthermore measures accurate intrinsic luminosities. Together these
allow us to determine the stellar Mass--Luminosity relation, and to do
it with more stars and over a wider mass range that has yet been
performed. (As Johannes Andersen pointed out after my talk, it may not
be useful to talk about achieving an accurate Mass--Luminosity
relation because such a thing does not exist, in the sense that the
luminosity of a star is not uniquely determined by its mass, but also
by abundance, rotation etc. Gaia could nonetheless shed
some light on these extra dependencies.)

\section{Spiral arms, ice ages and mass extinctions}
\vspace*{2ex}

Several papers have reported on a correlation and possible causal
connection between spiral arm passages and ice ages and/or mass
extinctions on the Earth.  One possible mechanism is the exposure to
massive stars in star forming regions: the increased cosmic ray flux
from type Ib/II supernovae within 10 pc during such passges could increase terrestrial
cloud cover (through water drop nucleation) and thus lower global
temperatures for millions of years (e.g. \cite{ellis95}), and/or the
OB star UV flux could destroy the ozone layer and cause widespread
extinction. A second mechanism is that the passage through regions of
larger stellar and gas density found in spiral arms could perturb the
Oort cloud and send minor bodies into the inner solar system, where
they could impact on the Earth.

Through accurate measurements of the 3D kinematics of sars, Gaia will
map the gravitational potential of the Galaxy and accurately determine
the velocity of the Sun.  Via numerical integration we can then
reconstruct the path of the Sun through the Galaxy over the last few
hundred million years and examine whether past ice ages and mass
exinctions coincide with spiral arms passages.  This has been done by
\cite{gies05} using the potential and solar motion derived from
Hipparcos by Dehnen \& Binney
(1998a,b)\cite[]{dehnen98a}\cite[]{dehnen98b}. While they found some
correlation between arm passages and ice ages, the results depended
heavily on the poorly known position and velocity (pattern speed) of
the spiral arms.

With Gaia we can dramatically improve this analysis: the astrometric
accuracies are better than Hipparcos by a factor of 500 (12-25\,\uas\
compared to 1000\,\uas), it includes many more stars (1 billion
compared to 120 000) and extends to fainter magnitudes (20 rather than
12.4).  With these data we can determine the gravitational potential
at higher spatial resolution and therefore reconstruct the solar
motion more accurately. Moreover, we can measure the position and
velocities of the spiral arms themselves from the Gaia observations of
their OB star population without assuming a rotation curve or needing
to know the extinction.  For an OB star at a distance of 4\,kpc from
the Sun observed through 4 magnitudes of extinction, Gaia will
determine its distance to an accuracy of 13\%, its space velocity to
about 1\, km/s.  Gaia can do this for some 50 000 OB stars within a
few kpc, allowing to build a more accurate model for the spiral arm
kinematics and thus make more conclusive statements about the
correlation between arm passages and Earth cataclysms.

Finally, Gaia will make a survey of Near-Earth Objects
(NEOs).  Gaia does real-time onboard object detection, so is sensitive
to transient phenomena and fast moving objects.  Gaia is predicted to detect some
16\,000 NEOs (\cite{mignard02}). While ground-based
surveys will discover many more in the coming years, Gaia can derive
accurate orbits and is sensitive to parts of the orbital parmeter
space which cannot easily be reached from the ground. It should
make a significant contribution to the census of potential
Earth impactors.

\end{document}